\documentclass[preprint]{aastex}
\usepackage{epsf}

\def\kms{\ifmmode {\, \rm km \, s^{-1}}\else {$\, \rm km \, s^{-1}$}\fi }
\def\rsun{\ifmmode {\rm R_{\odot}}\else $\rm R_{\odot}$\fi}
\def\msun{\ifmmode {\rm M_{\odot}}\else $\rm M_{\odot}$\fi}

\def\linebreak{\hfil\break}

\def\\{\hfil\linebreak}
\def\
{\hfil\linebreak}
\def\singlespace{\baselineskip=14pt}

\def\singlespace{%
    \lineskip                .15ex
    \baselineskip            3.0ex
   \lineskiplimit              0ex
   \parskip                0.60ex plus .30ex minus .15ex
   }%

\def\etal{{\it et al}. }
\def\msun{\ifmmode {\rm M_{\odot}}\else $\rm M_{\odot}$\fi}
\def\MSUN{\ifmmode {\rm M_{\odot}}\else $\rm M_{\odot}$\fi}
\def\msunyr{\ifmmode {\rm M_{\odot}\,yr^{-1}}\else $\rm 
M_{\odot}\,yr^{-1}$\fi}
\def\mdot{\ifmmode {\dot{M}}\else $\dot{M}$\fi}
\def\degree{\ifmmode {^\circ}\else {$^\circ$}\fi}
\def\mum{\ifmmode {\rm \mu {\rm m}}\else $\rm \mu {\rm m}$\fi}
\def\arcsec{\ifmmode ^{\prime \prime}\else $^{\prime \prime}$\fi}

\def\inch{\ifmmode ^{\prime \prime}\else $^{\prime \prime}$\fi}
\def\arcmin{\ifmmode ^{\prime}\else $^{\prime}$\fi}

\def\msun{\ifmmode {\rm M_{\odot}}\else $\rm M_{\odot}$\fi}
\def\lsun{\ifmmode {\rm L_{\odot}}\else $\rm L_{\odot}$\fi}
\def\mearth{\ifmmode {\rm M_{+\mskip-14.6muO\,}}\else $\rm 
M_{+\mskip-14.6muO\,}$\fi}
\def\mearth{\ifmmode {\rm M_{\earth}}\else $\rm M_{\earth}$\fi}
\newbox\grsign \setbox\grsign=\hbox{$>$} \newdimen\grdimen 
\grdimen=\ht\grsign
\newbox\simlessbox \newbox\simgreatbox
\setbox\simgreatbox=\hbox{\raise.5ex\hbox{$>$}\llap
     {\lower.5ex\hbox{$\sim$}}}\ht1=\grdimen\dp1=0pt
\setbox\simlessbox=\hbox{\raise.5ex\hbox{$<$}\llap
     {\lower.5ex\hbox{$\sim$}}}\ht2=\grdimen\dp2=0pt

\begin{document}

%\pagestyle{empty}

%\submitted{The Astrophysical Journal, submitted}
\title{The Eclipsing Binary BG Geminorum: Improved Constraints 
on the Orbit and the Structure of the Accretion Disk}
\author{Scott J. Kenyon, Paul J. Groot\footnote{present
address: Department of Astrophysics, University of Nijmegen, 
P.O. Box 9010, 6500 GL Nijmegen, The Netherlands}}
\affil{Smithsonian Astrophysical Observatory, 
60 Garden Street, Cambridge, MA 02138}
\author{and}
\author{Priscilla Benson}
\affil{Wellesley College, Whitin Observatory,
106 
Central Street, Wellesley, MA 02181-8286}
%\submitted{Received: 7 July 1999, Accepted: }
%
%\centerline{submitted to}
%\centerline{{\it The Astrophysical Journal Letters}}
%\centerline{March 1999}

%\singlespace

\begin{abstract}

We describe new optical photometric and spectroscopic observations 
of the semi-detached eclipsing binary BG Geminorum.  A large change 
in the amount of Mg I absorption at secondary maximum indicates the 
presence of cool material in the outer edge of the disk surrounding 
the unseen primary star.  Detection of weak He~I emission implies a 
hot radiation source at the inner edge of the disk.  If the velocity 
variations in the H$\beta$ emission line track the orbital motion of 
the primary star, the primary star has an orbital semiamplitude of 
$K_1$ = $K_{H \beta}$ = 16.0 $\pm$ 4.6 km s$^{-1}$. This result yields 
a mass ratio, $q = 0.22 \pm 0.07$, consistent with the $q = 0.1$ derived 
from the large ellipsoidal variation.  Despite this progress, the nature 
of the primary star -- B-type star or black hole -- remains uncertain.

\end{abstract}

\subjectheadings{binaries: eclipsing -- binaries: spectroscopic --
stars: emission-line -- stars: evolution -- stars: individual (BG Gem)}

\section{INTRODUCTION}

BG Geminorum was discovered by \citet{hof33} and \citet{jen38} as 
a possible RV Tau star with an uncertain
period of $\sim$ 60 days.  With a photographic magnitude of
$\sim$ 14, the star languished in the General Catalog of Variable
Stars \citep{kho85} until 1992, when \citet{ben00} discovered
an ellipsoidal variation with an optical amplitude of $\sim$
0.5 mag and a period of 91.645 days.  Light curves derived from
optical spectrophotometry reveal a deep primary eclipse at
short wavelengths, $\lambda \lesssim$ 4400 \AA, and a shallow
secondary eclipse at longer wavelengths.  The ellipsoidal light
curve and the radial velocity curve of the K0 I secondary star
indicate a semi-detached binary, where the K0 star transfers
material into an accretion disk around an unseen primary star.
The primary star is either an early B-type star or a black hole
\citep{ben00}.

This paper reports new optical and spectroscopic observations of BG Gem.
These data yield an orbit for H$\beta$ emission from the disk.  Coupled 
with an improved ephemeris and orbit for the K0 secondary, this result 
indicates a mass ratio, $ q \equiv M_{K0}/M_1$ = 0.22 $\pm$ 0.07,
where $M_1$ is the mass of the primary star and $M_{K0}$ is the mass
of the secondary star.  This result is consistent with the 
$q \approx$ 0.1 derived from the optical light curve.  The 
observations also provide additional information on the structure 
of the disk. A significant increase in Mg I absorption at secondary 
minimum shows that the outer part of the disk is cool, with 
$T \sim$ 4000 K at $R \approx$ 0.26 AU from the primary star.  
Despite the detection of weak He~I emission lines, the nature of the
central ionizing source is uncertain.  The absolute H~I and He~I line 
fluxes appear to rule out a B-type primary star as the ionizing source
and may indicate radiation from a boundary layer at the inner edge 
of the disk.

We begin with a summary of the observations in \S2, continue with 
the analysis in \S3, and conclude with a brief discussion in \S4.

\section{OBSERVATIONS}

Student observers acquired optical photometry of BG Gem with 
standard VR$_{\rm C}$I$_{\rm C}$ filters and a Photometrics
PM512 camera mounted on the Wellesley College 0.6-m Sawyer telescope.
\citet{ben00} describe the instrument and the data reduction.
Because Wellesley is not a photometric site, we derive 
photometry of BG Gem relative to two comparison stars.
The relative photometry has 1$\sigma$ probable errors of
$\pm$0.011 mag at V, $\pm$0.014 mag at R$_{\rm C}$, and
$\pm$0.011 mag at I$_{\rm C}$ \citep{ben00}.  Table 1 lists 
new relative photometry of BG Gem and a comparison star as a 
function of time (the Heliocentric Julian Date, JD) and
photometric phase ($\phi$).  These new data yield an improved 
ephemeris for primary minimum and photometric phase, 
\begin{equation}
\rm Min = JD~2449088.96 \pm 0.94 ~ + ~ (91.60 \pm 0.55) \cdot E ~ .
\end{equation}

P. Berlind, M. Calkins, and several other observers acquired
low resolution optical spectra of BG Gem with FAST, a high 
throughput, slit spectrograph mounted at the Fred L. Whipple 
Observatory 1.5-m telescope on Mount Hopkins, Arizona \citep{fab98}.
They used a 300 g mm$^{-1}$ grating  blazed at 4750 \AA, 
a 3\arcsec~slit, and a thinned Loral 
512 $\times$ 2688 CCD.  These spectra cover 3800--7500 \AA~at 
a resolution of 6 \AA.  We wavelength-calibrate the spectra 
in NOAO IRAF\footnote{IRAF is distributed by the National Optical 
Astronomy Observatory, which is operated by the Association of 
Universities for Research in Astronomy, Inc. under contract to the
National Science Foundation.}.
After trimming the CCD frames at each end of the slit, 
we correct for the bias level, flat-field each frame, 
apply an illumination correction, and derive a full wavelength 
solution from calibration lamps acquired immediately after each 
exposure.  The wavelength solution for each frame has a probable 
error of $\pm$0.5 \AA~or better.  To construct final 1-D spectra, 
we extract object and sky spectra using the optimal extraction 
algorithm APEXTRACT within IRAF.  Most of the resulting spectra 
have moderate signal-to-noise, S/N $\gtrsim$ 30 per pixel.

We derive radial velocities from the strong absorption and 
emission lines on FAST spectra.  For absorption line velocities, 
we cross-correlate the FAST spectra against the best-exposed 
spectrum, where the velocity is set by cross-correlation against 
standard stars with known velocities \citep[see][]{ton79,kur98}.
To avoid contamination from the hot primary,
we restrict the cross-correlation to $\lambda\lambda$5000--6800.
We measure emission line velocities from cross-correlations with
an emission-line template, as described by \citet{kur98}.
We adopt the velocity of H$\beta$ as the emission line velocity,
because H$\alpha$ may be blended with [N~II] emission on our
low resolution spectra.
We estimate errors of $\pm$30 \kms~ for absorption lines and 
$\pm$40 \kms~for emission lines \citep[see also][]{kur98}. 
Table 2 contains the first ten 
radial velocity measurements. The third column lists the measured 
absorption velocities as a function of JD and $\phi$; the fourth 
column lists measurements for H$\beta$.  The electronic version
of this paper includes all 189 radial velocity measurements.  

To analyze the phase variations of the absorption and emission line
spectra, we measure continuum magnitudes and indices using narrow 
passbands \citep{oco73,wor94}.  Table 3 of \citet{ben00} lists the 
central wavelength $\lambda$ and width $\delta \lambda$ for each. 
The absorption and emission indices, derived using SBANDS within IRAF, 
are $I_{\lambda}$ = $-$2.5 log($F_{\lambda}$/$\bar{F}$),
where $F_{\lambda}$ is the average flux in the passband $\lambda$ and
$\bar{F}$ is the continuum flux interpolated between the fluxes in
the neighboring blue band centered at $\lambda_b$ and the red band 
centered at $\lambda_r$.
Table 3 lists the first ten measured indices along with emission 
line equivalent widths as a function of JD and $\phi$.  The electronic
version of this paper contains all 191 measurements of absorption
indices and emission line equivalent widths.

To search for weak high ionization emission lines in BG Gem, we
examine coadded spectra.  We construct spectra in 20 phase bins with 
width $\Delta \phi$ = 0.05 at integral multiples of $\phi$ = 0.05.
The spectra provide fair detections of He I $\lambda$4471 and 
$\lambda$5876 at most orbital phases with equivalent widths of 
0.5 $\pm$ 0.2 \AA. The lines disappear at primary eclipse.  These data 
provide no evidence for higher ionization features such as He~II 
$\lambda$4686.  The coadded spectra suggest a weak broad pedestal 
of H$\alpha$ and H$\beta$ emission with a velocity width of 
$\sim$ 4000 km s$^{-1}$.  At low resolution, these features are 
difficult to disentangle from numerous K-type absorption features
and may not be real.

\section{ANALYSIS}

\subsection{Radial Velocities}

The absorption line radial velocities show clear evidence for orbital 
motion correlated with the photometric phase (Figure 1).  We analyze 
these observations using the \citet{mon79} Fourier transform algorithm 
\citep[see also][]{ken86}.  The best spectroscopic period,
$P_{spec} = 91.61 \pm 2.53$ days, agrees with the photometric
period.  A circular orbit with $P = P_{phot}$ from equation
(1) fits the orbit well.  This solution has an orbital 
semi-amplitude, $K_{\rm K0} = 71.7 \pm 3.4$ \kms,
and a fractional semi-major axis,
$A_{\rm K0}~{\rm sin}~i = 0.60 \pm 0.03$ AU.
Spectroscopic conjunctions occur 0.69 days prior to
photometric minima, with

\begin{equation}
\rm Conj = JD~2451012.28 \pm 0.71 + 91.60 ~ \cdot ~ E ~ .
\end{equation}

\noindent
The phase difference between primary eclipses and spectroscopic 
conjunctions has a significance of 1$\sigma$.
The mass function is 
$M_1^3 ~ {\rm sin}^3 i = (3.5 \pm 0.5)~\msun ~ (M_1 + M_{\rm K0})^2$.
If sin $i$ = 1, this result becomes
$M_1 \approx 3.5 ~ \msun ~ (1 + q)^2$,
where $q \equiv M_{\rm K0}/M_1$ is the mass ratio.
The eclipses place an upper limit on the mass of the primary star.
For $i$ $\gtrsim$ 65$^{\rm o}$, 
$M_1 \lesssim 5.7 ~ \msun$ for $q$ = 0.1 and
$M_1 \lesssim 7 ~ \msun$ for $q$ = 0.2.

Solutions with eccentric orbits may improve the fit to the 
data.  Iterations in Fourier and configuration space yield 
$e = 0.09 \pm 0.03$. The \citet{luc71} test suggests that
the non-zero $e$ is marginally significant.  We derive
Lucy-Sweeney probabilities of $p_{LS}$ = 3.6\% for the
Fourier solution and $p_{LS}$ = 4.9\% for the configuration
space solution; $p_{LS} <$ 5\% implies the eccentricity is 
non-zero with 95\% confidence. Radial velocity measurements 
using higher resolution spectra could resolve this ambiguity.
In this paper, we adopt the circular solution with the 
parameters quoted above.  If the K0~I secondary has negligible 
mass and sin $i$ = 1, these results place a lower limit on 
the mass of the primary, $M_1 \gtrsim$ 3.5 \msun.

Radial velocities of the H$\beta$ emission line also vary with
photometric phase (Figure 2).  The line variations during primary
minimum are consistent with the eclipse of a rapidly rotating disk, 
where the blue-shifted half of the disk is eclipsed first during 
ingress and is revealed first during egress.  The velocity 
variations of H$\beta$ outside primary minimum provide a 
direct estimate of the mass ratio.  The best-fitting circular 
orbit to data with $\phi$ = 0.05--0.95 has an
orbital semi-amplitude, $K_{H \beta}$ = $16.0 \pm 4.6$ km s$^{-1}$
and a fractional semi-major axis, $A_{H \beta}$ sin $i$ = $0.13 \pm 0.04$.
The orbital conjunction derived from the H$\beta$ velocities, 
Conj(H$\beta$) = JD 2451009.65 $\pm$ 4.6,
is 2.91 days prior to primary eclipse and
is 2.63 days prior to the spectroscopic conjunction defined
by the absorption line velocities.  Together with the 
absorption line orbit, these results indicate that the H$\beta$
line provides a reasonable first estimate for the orbit of the disk
surrounding the primary star.  If the orbital semi-amplitude
of the unseen primary is 16.0 km s$^{-1}$, the mass ratio is 
$q = 0.22 \pm 0.07$.  For sin $i$ $\approx$ 1, the component 
masses are then $M_1 = 4.3 \pm 0.8$ $M_{\odot}$ and
$M_{K0} = 0.95 \pm 0.45$ $M_{\odot}$.
These results are consistent with the $q \approx$ 0.1 derived
from the ellipsoidal light curve \citep{ben00}.

\subsection{Mg I Absorption from the Disk}

Most of the low excitation absorption lines in BG Gem spectra
provide a good measure of the spectral type for the K-type 
secondary star \citep{ben00}.  The Ba~I blend at 6495~\AA~is
constant with photometric phase (Figure 3). Absorption lines at 
shorter wavelengths vary consistently with photometric phase
(Figure 3; Benson et al. 2000).  All lines grow stronger at
$\phi$ = 0, when the K-type star eclipses light from the primary.
The absorption line strengths at primary minimum are consistent 
with a K0 I spectral type for the secondary star \citep{ben00}.

The equivalent widths of several absorption lines rise at 
$\phi \approx$ 0.5, when the disk surrounding the primary eclipses 
the secondary star.  The Na~I, Mg~I, and several other low ionization 
lines are as strong at secondary minimum as at primary minimum 
(Figure 3).  The disk surrounding the primary star is the only 
possible source of this extra absorption.  The duration of the 
second maximum in Mg I line strength yields a reliable estimate
for the sum of the radius of the K0 I secondary and the radius 
of the region that produces Mg~I absorption, 
$R_{\rm K0~I} + R_{\rm Mg~I}$ $\approx$ (0.60 $\pm$ 0.03) A;
this yields $R_{\rm K0~I} + R_{\rm Mg~I} \approx$ 0.44 AU for
$A$ = 0.73 AU.  With $R_{\rm K0~I} \approx$ 0.18 AU \citep{ben00},
the size of the Mg~I region is then $R_{\rm Mg~I}$ $\approx$ 0.26 AU.  
This radius is larger than the $R_{\rm H \alpha} \approx$ 0.22 AU 
derived from the emission line eclipses \citep{ben00}.

The Mg~I line strength appears to drop rapidly at $\phi$ = 0.5, when 
our line-of-sight passes through the center of the primary and the
surrounding accretion disk.  The line strength recovers at $\phi$ = 
0.51 and then falls gradually as the system approaches orbital 
quadrature.  Although our phase coverage at $\phi$ = 0.5 is not
superb, this behavior repeats over several eclipses.

We propose a simple explanation for the behavior of Mg~I absorption
at $\phi$ = 0.45--0.55.  Because Mg~I has an ionization potential
of 7.6 eV, Mg is ionized to Mg~II in the inner region of the disk 
that produces H$\alpha$ emission.  If the outer part of the disk is 
cool, T $\sim$ 4000 K, most Mg atoms are neutral and can absorb radiation 
from the K0 I secondary.  We see this absorption at secondary minima 
when the K0 star lies behind the disk. The drop in Mg~I absorption 
at $\phi$ = 0.5 implies a ring geometry for Mg I atoms.  The central 
hole of the ring, where magnesium is Mg~II, yields a longer path 
length through the edge of the ring than through the center.  

If the outer disk is heated by the central star, the Mg I absorption
places some useful limits on the temperature of the central star.
In most heating models, the disk temperature varies with radius
as $T_d \propto R^{-n}$ with $n \approx \frac{1}{2}$ to $\frac{3}{4}$
\citep{kh87}. With $T_d \approx$ 4000 K at $R \approx$ 0.26 AU,
the limits on the temperature of the 4--5 $M_{\odot}$ main sequence 
star are reasonable, $T_{\star} \sim$ 12,000~K to 20,000~K.

\subsection{He I Emission}

We detect modest He~I emission on coadded spectra of BG Gem.  The
$\lambda$5876 line is blended with strong Na~I D absorption lines
and is difficult to measure accurately.  The $\lambda$4471 line has
an equivalent width of $\sim$ 0.5 \AA~at orbital quadratures and 
$\sim$ 1 \AA~at $\phi$ = 0.5.  Both lines disappear at primary eclipse.
The integrated line flux is $F_{4471} \approx$ $7 \times 10^{-14}$ 
erg cm$^{-2}$ s$^{-1}$ for a visual reddening $A_V$ = 1.65 mag \citep{ben00}. 
The total $\lambda$4471 luminosity is $L_{4471} \approx $ 
$5 \times 10^{31}$ erg s$^{-1}$ (d/2.5 kpc)$^2$.  The He~I 
$\lambda$5876 luminosity is roughly a factor of two larger, but 
is more uncertain due to blending with Na~I~D.  No other obvious 
He~I or He~II lines are visible on the coadded spectra.

Detection of He~I emission implies a luminous source of high energy 
photons.  For case B recombination, the observed $F_{4471}/F_{H \beta}$
intensity ratio requires an ionizing source with an effective temperature
$T_{eff} \approx$ 25,000 K \citep{ost88}.  If the atmosphere of the 
primary star produces these photons, the observed H$\beta$ and He~I 
$\lambda$4471 fluxes imply a B1--B2 V star with $L_1$ $\sim$ 5000 
$L_{\odot}$ \citep{spi78,sha97}.  Normal B1--B2 V stars have masses 
of 15--20 $M_{\odot}$ \cite[e.g.,][and references therein]{vac96,sha97}.
These results are not consistent with a 4 $M_{\odot}$ B5 main sequence 
star with $L \approx$ 1000 $L_{\odot}$.  The line fluxes thus preclude 
photoionization from a normal B-type star.

The dynamical mass and emission line fluxes are also inconsistent with 
a Be-type primary star.  Known Be stars with He~I emission have B1--B2 
or earlier spectral types; Be stars with B4--B6 spectral types have 
weak He~I absorption lines \citep{jas87}.  Known Be stars have very steep 
Balmer decrements indicative of optically thick H$\alpha$; H$\delta$, 
H$\epsilon$, and higher level transitions in the series are almost always 
strong absorption lines \citep{jas87,tor97}.  We detect H8 and H9 emission
in BG Gem; the line fluxes are consistent with case B recombination 
for $A_V$ = 1.65 mag \citep{ben00}.  Be-type stars with strong H$\alpha$ 
emission also have large near-IR excesses from dusty circumstellar 
material \citep{dou94}. Near-IR photometry demonstrates that BG Gem has no
substantial near-IR excess \citep{ben00}. Thus, the primary star in BG Gem 
is not a normal Be-type star.

Accretion provides another energy source for the H~I and He~I emission.
The double-peaked H$\alpha$ and H$\beta$ lines form in the disk 
surrounding the primary star.  The inner part of this disk produces 
blue continuum emission which dominates the spectrum for 
$\lambda \lesssim$ 4500 \AA.  There are two possible sources of 
high energy photons, the inner disk and the boundary layer between 
the inner disk and the stellar photosphere.  If the inner disk has 
a temperature of $\sim$ 25,000 K, the boundary layer temperature should 
exceed $T_{bl} \sim 10^5$ K \citep{lyn74,ken84}. The lack of high 
ionization emission lines -- as observed in some symbiotic stars 
\citep{ken84,ken91} -- precludes such a hot boundary layer. If instead 
the boundary layer alone is responsible for the He~I emission, the radius 
of the central star is roughly \citep{ken91}
\begin{equation}
R_1 \approx 5.5 ~ R_{\odot} \left ( \frac{L_{bl}}{1000 ~ L_{\odot}} \right )^{0.4} \left ( \frac{T_{bl}}{25,000 ~ {\rm K}} \right )^{-1.8} ~ . 
\end{equation}
\noindent
For a boundary layer which radiates as a star with $T_{bl}$ = 25,000 K,
the bounary layer luminosity is $L_{bl} \approx$ 3000 $L_{\odot}$.
The stellar radius is $R_1 \approx$ 8.5 $R_{\odot}$.  This solution 
implies an evolved primary star with $L_1 \approx$ 3500 $L_{\odot}$,
$T_{eff} \approx$ 15,000 K, and a dynamical mass of $\sim$ 4 $M_{\odot}$.  
These properties are consistent with those derived from Mg~I absorption 
lines produced in the outer disk.
Because we do not detect optical spectral features from this star,
this solution is viable only if the outer edge of the disk hides the
star from view.  The accretion rate in this interpretation is large,
$\dot{M} \approx$ 4--5 $\times 10^{-4} M_{\odot}$ yr$^{-1}$.  This 
large accretion rate implies very bright optical emission from the
disk \citep{ken84} which is inconsistent with the optical spectrum of
BG Gem unless $i \gtrsim$ 86$^{\rm o}$.

\section{DISCUSSION AND SUMMARY}

Our new optical spectra yield a better understanding of the orbit, geometry,
and disk structure in BG Gem.  Absorption line radial velocities derived 
from these spectra reduce the uncertainties in the orbital parameters by 
$\sim$ 25\% and confirm that the minimum mass for the primary star is 
3.5--4.0 $M_{\odot}$.  Using the H$\beta$ emission line, we also derive 
a tentative orbit for the disk surrounding the primary star.  Although 
the interpretation of orbits derived from material in a disk is 
controversial \citep[e.g.,][]{war95}, the mass ratio $q \approx$ 0.2
inferred from this orbit is consistent with the $q \approx$ 0.1 derived 
from the optical light curve \citep{ben00}.  The velocity amplitude of 
$K_{H \beta}$ = $16.0 \pm 4.6$ km s$^{-1}$ is at the limit of our FAST 
spectra; improvements in this measurement require higher resolution spectra.

The behavior of Mg I and other neutral absorption lines at $\phi \approx$ 
0.5 implies a cool ring of material in the outer disk with $T \sim$ 4000 K. 
The outer radius of this ring, $\sim$ 0.26 AU, is slightly larger than the 
radius of the ionized region that produces H~I emission.  The outer radius 
of the Mg I ring is $\sim$ 70\% of the radius of the Roche lobe of the
primary star, well within the boundary where tidal forces truncate the disk.

The detection of weak He~I emission implies a modest source of high energy
photons from the primary star, the inner disk, or the boundary layer between
the inner disk and the primary star.  The ratio of the He~I $\lambda$4471
flux to the H$\beta$ flux implies a temperature of $\sim$ 25,000 K for
the ionizing source.  If the disk does not occult any of the ionized 
H~II region, the luminosity of the ionizing source is $\sim$ 1000--5000 
$L_{\odot}$.  If the primary is a B-type star, the dynamical mass 
estimates favor a B4--B6 giant which lies hidden behind the outer edge 
of the accretion disk.  For an orbital inclination $i \approx$ 85\degree, 
this picture implies a disk luminosity which exceeds the observed disk 
luminosity.  If the orbital inclination is larger than 85\degree, the 
outer disk probably hides some of the He~I and other moderate ionization 
emission lines from our line-of-sight.  In this case, we cannot make 
a meaningful derivation of the physical properties of the primary star.

Our data do not rule out the possibility that the primary star is a black 
hole \citep{ben00}.  Although we do not detect X-rays or high ionization 
emission lines -- such as the He~II features observed in other black hole 
binaries -- the outer edge of the disk can hide these features from our 
line-of-sight
for $i \gtrsim$ 85\degree~\citep{ben00}.  Our coadded spectra provide weak 
evidence for H$\alpha$ emission with velocities of $\sim$ 2000 km s$^{-1}$.
The maximum rotational velocity for material in a disk surrounding a B-type
star is $\sim$ 700 km s$^{-1}$.  If confirmed with higher resolution optical 
spectra, high velocity H$\alpha$ emission would preclude a B-type primary
star and favor a black hole primary star in BG Gem.

Making further progress on this system requires ultraviolet spectra and
more intensive analysis of the optical spectra.  We have an approved program
with the {\it Hubble Space Telescope} to acquire low resolution ultraviolet
spectra at specific orbital phases.  Detection of a strong ultraviolet
continuum or ultraviolet absorption lines from Si~II and Si~III (among
others) would indicate a B-type primary; failure to detect these features
or detection of (broad) high ionization emission lines would favor a
black hole primary.  Optical spectra acquired during primary eclipse 
provide an opportunity to use maximum entropy techniques to construct 
the surface brightness profile and eclipse maps of the accretion disk
surrounding the primary star \citep[e.g.,][]{hor85,mar88,war95,gro01,dia01}.
The cycle-to-cycle stability of BG Gem (see Figure 4) should yield good 
disk `images' to test for the presence of high velocity material in the 
disk and to search for the source of high energy photons at the inner 
disk.  These techniques might also clarify the geometry of Mg I absorption
in the outer disk. We plan to address these issues in future papers.

\vskip 6ex

We thank P. Berlind, M. Calkins, and other queue observers 
at the FLWO 60\inch~telescope for acquiring the FAST spectra 
used in this project. Susan Tokarz made the preliminary 
reductions of the FAST spectra.  PJG was supported by a CfA 
fellowship.

\vfill
\eject

\begin{center}
\begin{deluxetable}{l c c c c c c c}
\singlespace
\scriptsize
\tablenum{1}
\tablecaption{Optical Photometry}
\tablehead{
\colhead{JD} & \colhead{Phase} &
\colhead{$\delta$V(BG)} & \colhead{$\delta$V(com)} &
\colhead{$\delta \rm R_C$(BG)} & \colhead{$\delta \rm R_C$(com)} &
\colhead{$\delta \rm I_C$(BG)} & \colhead{$\delta \rm I_C$(com)}}
\startdata
48728.5230 & 0.065 & \nodata & \nodata &  $+$0.200 &  $+$0.369 & \nodata & \nodata \\
49059.5030 & 0.678 & $+$0.291 &  1.079 & \nodata & \nodata & \nodata & \nodata \\
50156.5270 & 0.655 & $+$0.232 &  1.044 & \nodata & \nodata & \nodata & \nodata \\
50168.4870 & 0.785 & \nodata & \nodata & \nodata & \nodata & $+$0.190 &  $+$0.682 \\
50500.5570 & 0.410 & \nodata & \nodata & \nodata & \nodata & \nodata & \nodata \\
50851.5620 & 0.242 & \nodata & \nodata & \nodata & \nodata & $-$0.202 &  $+$0.673 \\
50911.5113 & 0.897 & $+$0.343 & $+$1.091 &  $+$0.146 &  $+$0.428 & $-$0.057 &  $+$0.652 \\
51215.5767 & 0.216 & \nodata & $+$1.067 & $-$0.223 &  $+$0.382 & \nodata &  $+$0.705 \\
51276.5300 & 0.882 & \nodata & \nodata &  $+$0.122 &  $+$0.367 & $-$0.102 &  $+$0.720 \\
51577.5313 & 0.168 & $+$0.158 & $+$0.978 &  $+$0.003 &  $+$0.326 & $-$0.151 &  $+$0.679 \\
51580.5350 & 0.201 & \nodata & \nodata &  $+$0.025 &  $+$0.366 & \nodata & \nodata \\
51590.6033 & 0.311 & $+$0.171 & $+$1.060 &  $+$0.007 &  $+$0.346 & $-$0.160 &  $+$0.684 \\
51592.6013 & 0.332 & $+$0.220 & $+$1.053 &  $+$0.077 &  $+$0.367 & $-$0.126 &  $+$0.689 \\
51604.6123 & 0.463 & $+$0.734 & $+$1.041 &  $+$0.563 &  $+$0.362 &  $+$0.431 &  $+$0.709 \\
51609.5610 & 0.517 & \nodata & \nodata & \nodata & \nodata &  $+$0.408 &  $+$0.708 \\
51613.5735 & 0.561 & \nodata & \nodata &  $+$0.213 &  $+$0.440 &  $+$0.227 &  $+$0.607 \\
51627.6090 & 0.715 & \nodata & \nodata & \nodata & \nodata & $-$0.244 &  $+$0.798 \\
51952.6777 & 0.263 & $+$0.098 & $+$1.043 & $-$0.042 &  $+$0.365 & $-$0.264 &  $+$0.649 \\
51953.5237 & 0.273 & $+$0.050 & $+$0.965 &  $+$0.048 &  $+$0.449 & $-$0.156 &  $+$0.734 \\
51954.5090 & 0.283 & $+$0.129 & $+$1.063 &  $+$0.028 &  $+$0.403 & \nodata & \nodata \\
51958.6247 & 0.328 & $+$0.199 & $+$1.042 &  $+$0.043 &  $+$0.368 & $-$0.143 &  $+$0.697 \\
51959.5327 & 0.338 & $+$0.216 & $+$1.039 &  $+$0.042 &  $+$0.357 & $-$0.106 &  $+$0.700 \\
51962.6210 & 0.372 & $+$0.260 & $+$1.019 &  $+$0.123 &  $+$0.370 & $-$0.053 &  $+$0.695 \\
51967.5167 & 0.425 & $+$0.582 & $+$1.005 &  $+$0.398 &  $+$0.328 &  $+$0.059 &  $+$0.612 \\
51968.5760 & 0.437 & $+$0.482 & $+$0.986 &  $+$0.236 &  $+$0.328 & \nodata & \nodata \\
51993.5780 & 0.710 & $+$0.174 & $+$1.056 &  $+$0.006 &  $+$0.360 & $-$0.175 &  $+$0.686 \\
51996.5537 & 0.742 & $+$0.180 & $+$1.047 & $-$0.045 &  $+$0.327 & $-$0.207 &  $+$0.703 \\
51997.5813 & 0.754 & $+$0.149 & $+$1.038 & $-$0.020 &  $+$0.351 & $-$0.193 &  $+$0.693 \\
52004.6123 & 0.830 & $+$0.201 & $+$1.044 &  $+$0.038 &  $+$0.375 & $-$0.141 &  $+$0.681 \\
52005.5177 & 0.840 & $+$0.236 & $+$1.042 &  $+$0.083 &  $+$0.356 & $-$0.109 &  $+$0.679 \\
\enddata
\end{deluxetable}
\end{center}

\clearpage

\begin{center}
\begin{deluxetable}{l c r r}
\singlespace
\footnotesize
\tablewidth{0pt}
\tablenum{2}
\tablecaption{Radial Velocities$^a$}
\tablehead{
\colhead{JD} & \colhead{Phase} &
\colhead{$v_{abs}$} & \colhead{$v_{\rm H \beta}$}}
\startdata
51547.8309 & 0.844 & $-$16.790 & $+$26.140 \\
51547.8381 & 0.844 & $-$12.160 & $+$34.710 \\
51550.8141 & 0.876 & $+$11.680 & $+$57.490 \\
51552.7494 & 0.897 & $-$36.480 & $-$4.600 \\
51552.7524 & 0.897 & $-$42.240 & $-$15.880 \\
51555.8512 & 0.931 & $-$29.990 & $-$0.380 \\
51557.7808 & 0.952 & $+$32.920 & $+$118.580 \\
51571.6108 & 0.103 & $+$43.590 & $+$4.460 \\
51571.6147 & 0.103 & $+$39.510 & $+$7.680 \\
51572.7773 & 0.116 & $+$58.040 & $+$27.610 \\
\enddata
\tablenotetext{a} {The velocities are measured in km s$^{-1}$.}
\end{deluxetable}
\end{center}

\clearpage

\begin{center} 
\begin{deluxetable}{l c c c c c c c} 
\singlespace 
\scriptsize 
\tablewidth{0pt}
\tablenum{3} 
\tablecaption{Absorption and Emission Line Indices } 
\tablehead{ 
\colhead{JD} & \colhead{Phase} &  
\colhead{$I_{\rm Fe~I}$} & \colhead{$I_{\rm Mg~I}$} & 
\colhead{$I_{\rm Na~I}$} & \colhead{$I_{\rm Ba~I}$} & 
\colhead{EW$(H\beta)$} & \colhead{EW$(H\alpha)$} } 
\startdata 
51547.8309 & 0.844 & 0.10 & 0.13 & 0.15 & 0.05 & $-$8.900 & $-$30.580 \\
51547.8381 & 0.844 & 0.13 & 0.15 & 0.15 & 0.05 & $-$10.270 & $-$31.530 \\
51550.8141 & 0.876 & 0.13 & 0.14 & 0.15 & 0.04 & $-$10.090 & $-$31.320 \\
51552.7494 & 0.897 & 0.11 & 0.15 & 0.15 & 0.05 & $-$9.560 & $-$27.370 \\
51552.7524 & 0.897 & 0.10 & 0.15 & 0.15 & 0.05 & $-$6.320 & $-$22.750 \\
51555.8512 & 0.931 & 0.13 & 0.17 & 0.16 & 0.05 & $-$9.270 & $-$28.880 \\
51557.7808 & 0.952 & 0.14 & 0.17 & 0.15 & 0.05 & $-$9.260 & $-$28.950 \\
51571.6108 & 0.103 & 0.12 & 0.15 & 0.16 & 0.04 & $-$9.120 & $-$28.880 \\
51571.6147 & 0.103 & 0.11 & 0.15 & 0.15 & 0.05 & $-$9.200 & $-$29.160 \\
51572.7773 & 0.116 & 0.10 & 0.14 & 0.16 & 0.05 & $-$8.980 & $-$29.800 \\
\enddata
\end{deluxetable}
\end{center}

\clearpage

%\centerline{Figure Captions}

\hskip -5ex
\epsffile{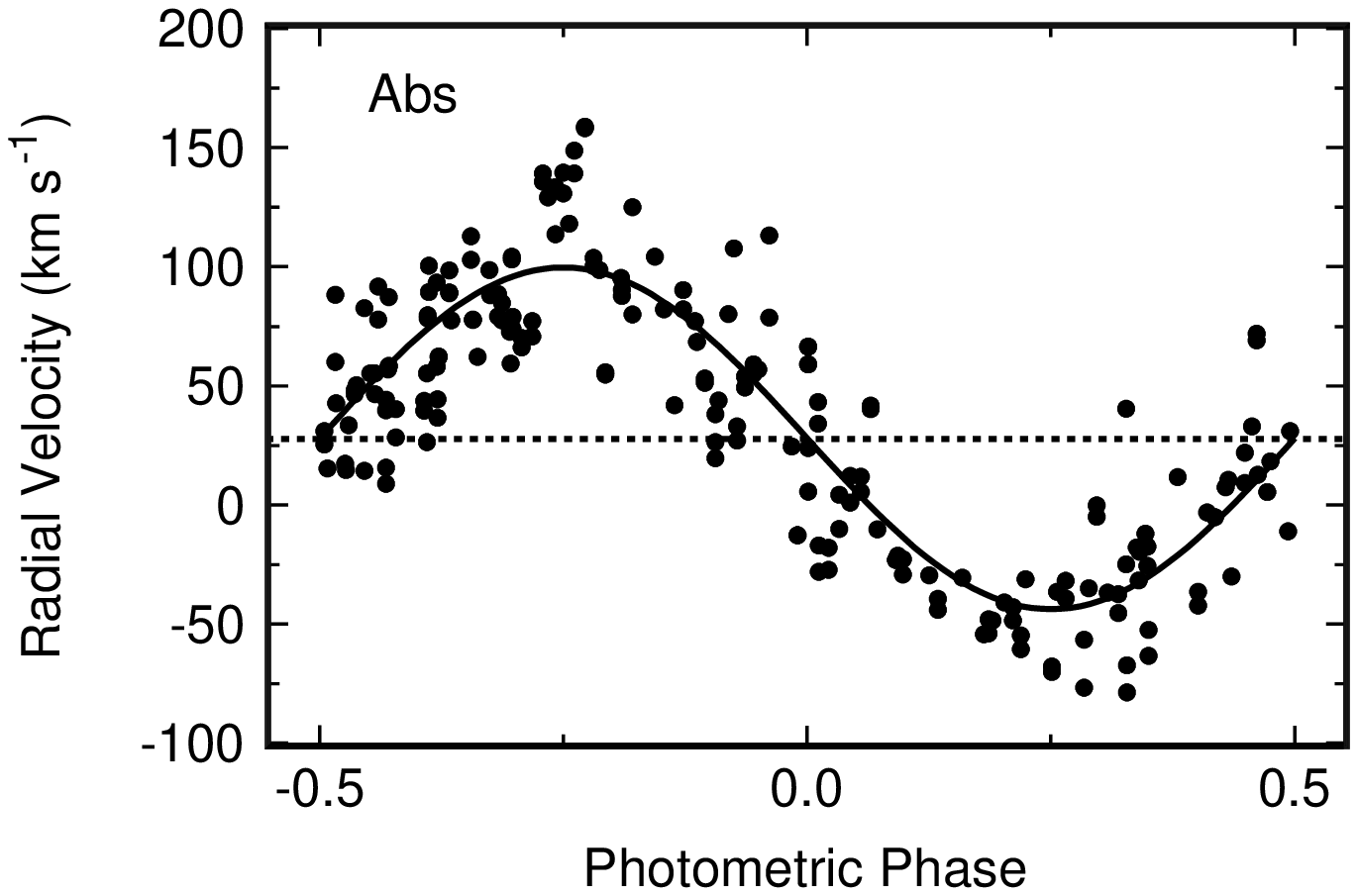}
\figcaption{Absorption line radial velocity curve.
The solid line is the best-fitting circular orbit to
the measured velocities (filled circles). The dashed
line indicates the systemic velocity.}  

\hskip -3ex
\epsffile{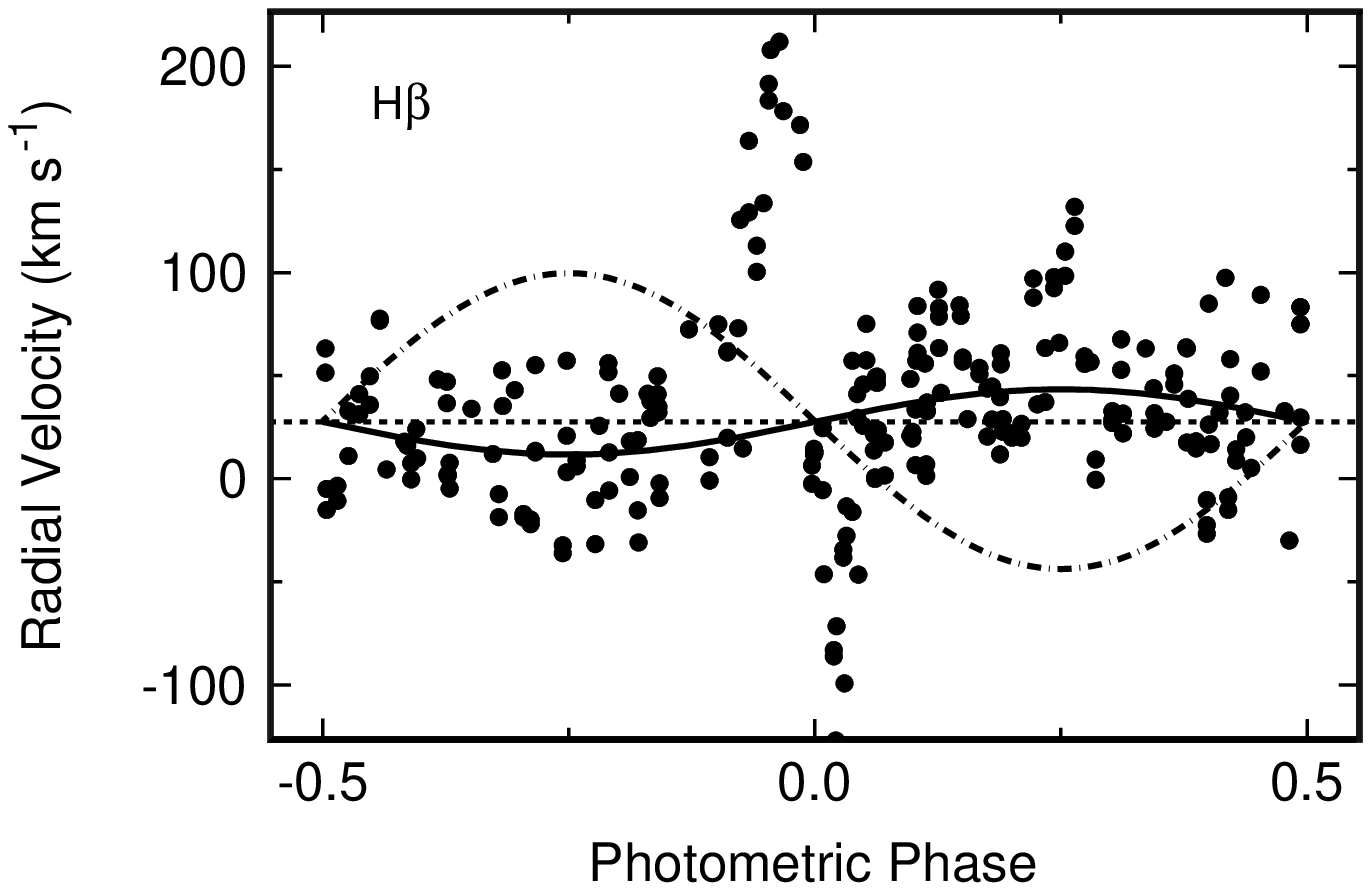}
\figcaption{Emission line radial velocity data for H$\beta$.
The dot-dashed curve repeats the absorption line orbit from
Figure 1; the dashed line indicates the systemic velocity
derived from the absorption line orbit. The solid line is
the orbit inferred from the H$\beta$ velocities outside of
primary eclipse ($\phi$ = 0.05--0.95).}

\hskip -1ex 
\epsffile{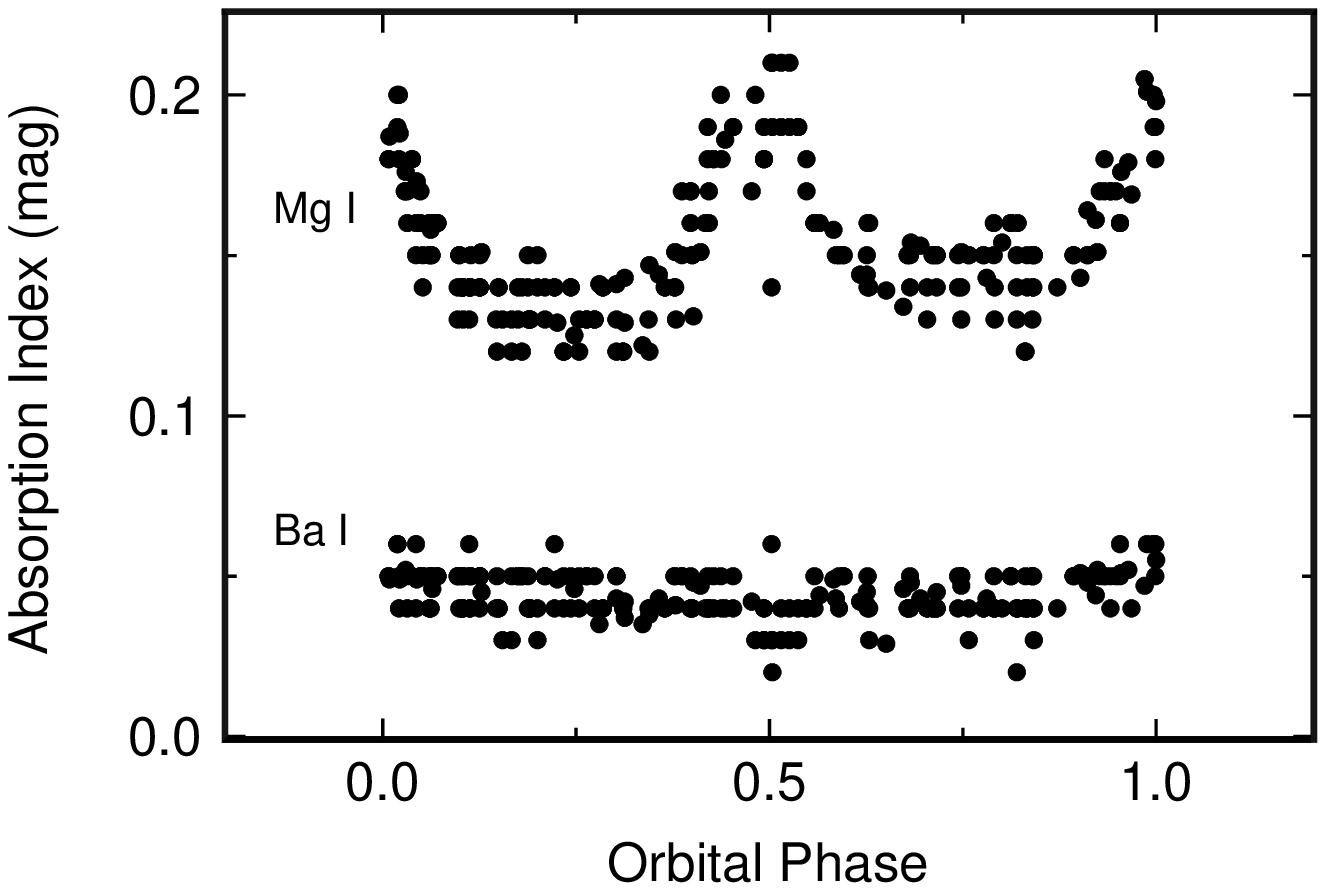}
\figcaption{Variation of absorption line indices. The Ba~I index
at 6495 \AA~is nearly constant with phase. The Mg~I index 
strengthens during primary eclipse and during secondary eclipse.}

\hskip -1ex 
\epsffile{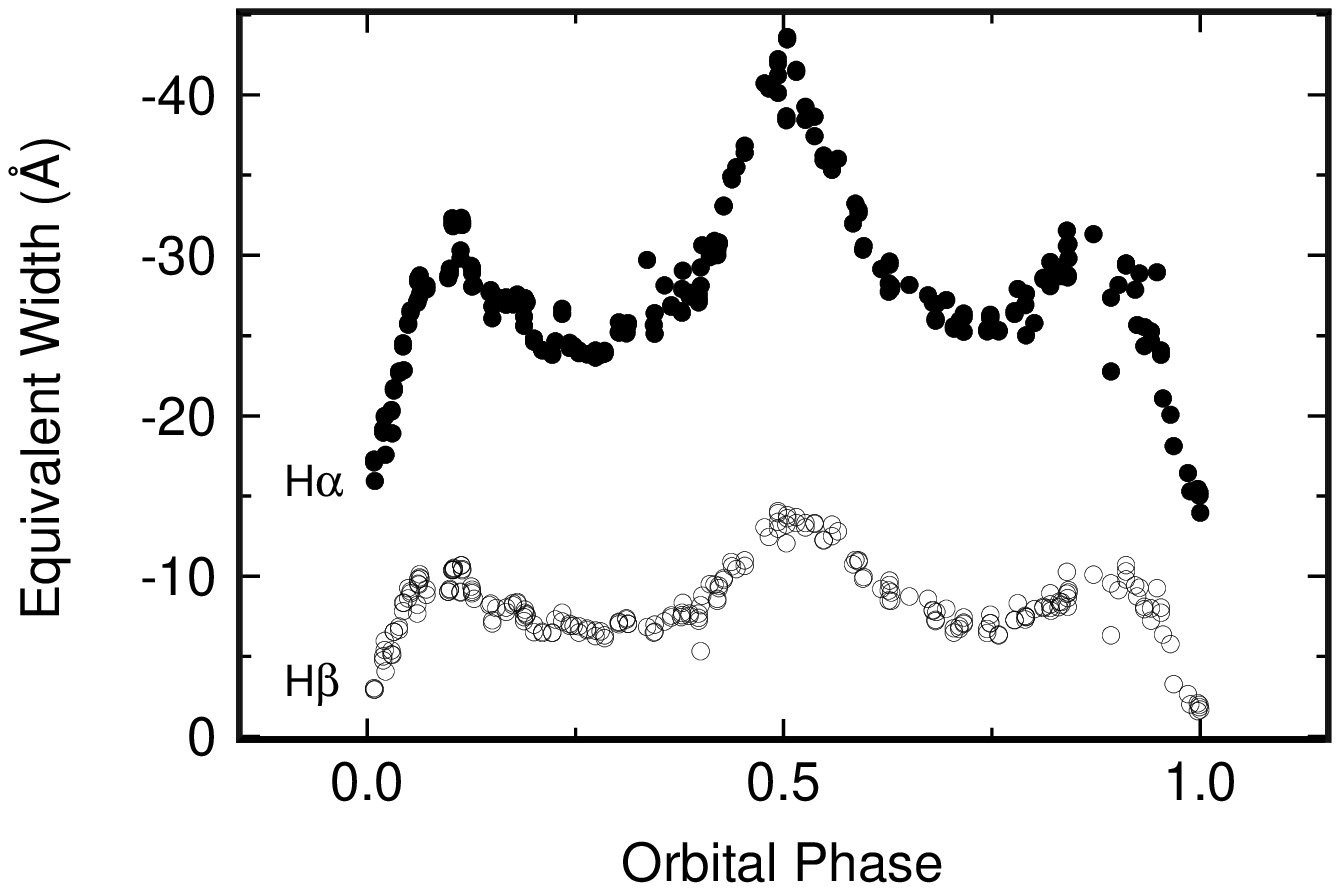}
\figcaption{Variation of emission line equivalent widths. 
The H~I emission is eclipsed by the secondary at $\phi$ = 0.
Because the continuum declines when the disk eclipses the K0 I
secondary star, the equivalent widths rise at $\phi$ = 0.5.
Outside of eclipse, the equivalent widths vary sinusoidally in 
phase with the ellipsoidal variations of the secondary.}

\end{document}